# Tip-Enhanced Infrared Difference-Nanospectroscopy of the Proton Pump Activity of Bacteriorhodopsin in Single Purple Membrane Patches

Valeria Giliberti,*,† Raffaella Polito,‡ Eglof Ritter,° Matthias Broser,° Peter Hegemann,° Ljiljana Puskar,∥ Ulrich Schade,∥ Laura Zanetti-Polzi,⊥ Isabella Daidone,⊥ Stefano Corni,#,$ Francesco Rusconi,± Paolo Biagioni,± Leonetta Baldassarre,‡ and Michele Ortolani*,†,‡

†Istituto Italiano di Tecnologia, Center for Life NanoScience, Viale Regina Elena 291, I-00161 Roma, Italy
‡Department of Physics, Sapienza University of Rome, Piazzale Aldo Moro 2, I-00185 Roma, Italy
°Humboldt-Universität zu Berlin, Institut für Biologie, Invalidenstraße 42, D-10115 Berlin, Germany
∥Helmholtz-Zentrum Berlin für Materialien und Energie GmbH, Albert-Einstein-Str. 15, 12489 Berlin, Germany
⊥Department of Physical and Chemical Sciences, University of L'Aquila, Via Vetoio, I-67010 L'Aquila, Italy
#Department of Chemical Sciences, University of Padova, Via Marzolo 1, I-35131 Padova, Italy
$CNR Institute of Nanoscience, Via Campi 213/A, I-41125 Modena, Italy
±Dipartimento di Fisica, Politecnico di Milano, Piazza Leonardo da Vinci 32, I-20133 Milano, Italy

**Ⓢ** Supporting Information

**ABSTRACT:** Photosensitive proteins embedded in the cell membrane (about 5 nm thickness) act as photoactivated proton pumps, ion gates, enzymes, or more generally, as initiators of stimuli for the cell activity. They are composed of a protein backbone and a covalently bound cofactor (e.g. the retinal chromophore in bacteriorhodopsin (BR), channelrhodopsin, and other opsins). The light-induced conformational changes of both the cofactor and the protein are at the basis of the physiological functions of photosensitive proteins. Despite the dramatic development of microscopy techniques, investigating conformational changes of proteins at the membrane monolayer level is still a big challenge. Techniques based on atomic force microscopy (AFM) can detect electric currents through protein monolayers and even molecular binding forces in single-protein molecules but not the conformational changes. For the latter, Fourier-transform infrared spectroscopy (FTIR) using difference-spectroscopy mode is typically employed, but it is performed on macroscopic liquid suspensions or thick films containing large amounts of purified photosensitive proteins. In this work, we develop AFM-assisted, tip-enhanced infrared difference-nanospectroscopy to investigate light-induced conformational changes of the bacteriorhodopsin mutant D96N in single submicrometric native purple membrane patches. We obtain a significant improvement compared with the signal-to-noise ratio of standard IR nanospectroscopy techniques by exploiting the field enhancement in the plasmonic nanogap that forms between a gold-coated AFM probe tip and an ultraflat gold surface, as further supported by electromagnetic and thermal simulations. IR difference-spectra in the 1450−1800 cm$^{-1}$ range are recorded from individual patches as thin as 10 nm, with a diameter of less than 500 nm, well beyond the diffraction limit for FTIR microspectroscopy. We find clear spectroscopic evidence of a branching of the photocycle for BR molecules in direct contact with the gold surfaces, with equal amounts of proteins either following the standard proton-pump photocycle or being trapped in an intermediate state not directly contributing to light-induced proton transport. Our results are particularly relevant for BR-based optoelectronic and energy-harvesting devices, where BR molecular monolayers are put in contact with metal surfaces, and, more generally, for AFM-based IR spectroscopy studies of conformational changes of proteins embedded in intrinsically heterogeneous native cell membranes.

**KEYWORDS:** Infrared spectroscopy, bacteriorhodopsin, transmembrane proteins, protein conformational changes, AFM-IR, plasmonic nanogap

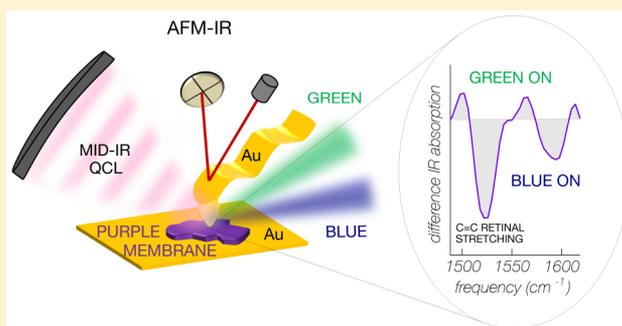

R hodopsins as photosensitive transmembrane proteins (TMPs)[1] are fundamental biological macromolecules that can act as light-gated ion pumps, channels, and enzymes that transport ions or information across the cell membrane (about 5 nm thickness), or as light receptors initiating enzymatic processes as phototaxis or vision.[2,3] They generally consist of seven transmembrane alpha-helices (opsin)[1] and a covalently bound photosensitive retinal cofactor that crucially absorbs light. Light absorption is followed by a trans to cis photoisomerization of the retinal.[4] The visible-light-absorption properties of the rhodopsins are mostly determined by specific interaction of retinal with residues of the retinal binding pocket forming the effective rhodopsin chromophore. Conformational changes of this chromophore and of the protein backbone are the key determinants of the photosensitive rhodopsin functions. In the relevant case of photoactivated pumps and channels, these conformational changes of the protein backbone lead to an opening of the conducting pore through which ions are then transported along the electrochemical gradient[5,6] or to p$K$-changes of amino acid residues that result in an uphill proton transport against the electrochemical gradient.[7,8]

Infrared (IR) spectroscopy is one of the main tools used for the investigation of the conformational changes of rhodopsins due to the high sensitivity of the IR vibrational absorption frequencies and lineshapes to structural features.[9,10] Moreover, IR spectroscopy can be applied to rhodopsins without interfering with the visible light employed for the optical activation of the membrane functionalities.[11] Because of the extremely small variations in the protein structure deriving from light-induced conformational changes of TMPs, FTIR spectrometers are usually employed in difference-spectroscopy mode, where IR vibrational absorption spectra are acquired with and without visible-light illumination, and often only the difference-spectrum is analyzed.[12] FTIR difference-spectroscopy is widely applied to the study of, for example, the light-gated proton pump bacteriorhodopsin (BR)[13] or the light-gated channelrhodopsin.[10,14] FTIR spectroscopy, however, lacks the spatial resolution and the sensitivity required to address individual parts of membranes and is therefore limited to highly concentrated suspensions or thick films containing about 10$^9$ cell membrane patches with a high density of light-sensitive TMPs, preferentially of one single species.[12] Efforts have been undertaken to increase the sensitivity of FTIR-based approaches by exploiting the attenuated total reflectance (ATR) effect[15] or, to reach down to the membrane monolayer level, by applying the surface-enhanced IR (SEIRA) absorption[16] by metal nanostructures.[17,18] The typical diameter of a native cell membrane patch is indeed of the order of 1 $\mu$m only and is not even addressable by FTIR microscopy (micro-FTIR).[19] Large-area TMP monolayers suitable for ATR-FTIR and SEIRA-FTIR difference-spectroscopy have been obtained by reforming self-assembled thin films where TMPs are embedded into an artificially reconstructed cell membrane.[18] However, this approach requires complex protein purification processes[20] and prevents the study of the TMPs in their native membrane environment.[21] More recently, the requirement of large sample areas of many squared millimeters covered by membrane monolayers, which represents a challenge for present TMP biotechnologies,[18,20] has been relaxed with the introduction of SEIRA structures in micro-FTIR, but signal-to-noise ratios are still to be improved.[22]

The atomic force microscope (AFM) is the instrument of choice for studying single native membrane patches and different AFM-based approaches have been applied to the study of photosensitive TMPs.[21,23−27] High-speed AFM[24] and conductive AFM[25,26] have been applied to probe the light-activated functions of BR embedded in 5 nm-thick individual patches of its native cell membrane, usually referred to as the purple membrane. Conductive AFM may, indeed, provide insights into local charge transport through BR molecules,[25,26] while high-speed AFM has been applied to shed light onto the main morphological changes of the BR backbone through fine topography variations.[24] These efforts are motivated by the potential use of BR as biomaterial for optoelectronic applications including organic photovoltaic cells,[28,29] where purple-membrane patches are deposited on flat metal electrodes. In particular, it has been found[30] that the photoelectrical activity of BR in the case of membrane monolayers on metal surfaces is much less efficient than the ideal mechanism in liquid suspensions or multilayers. However, a fundamental understanding of such decreased activity, and therefore of the underlying modifications of the BR photocycle imposed by the combination of several effects such as direct physisorption, chemisorption, or adhesion of BR on metal layers, is still lacking. One of the reasons is that a purely AFM-based technique cannot detect subtle conformational changes connected to proton transport in the inner protein structure. It is therefore desirable to combine the IR spectroscopy capabilities with AFM-based techniques as already proposed by many authors;[31−34] however, it has not yet been applied to IR difference-spectroscopy of TMPs.

In this work, we present the first IR difference-nanospectroscopy measurement of conformational changes of a photosensitive TMP using the bacteriorhodopsin mutant BR D96N,[35] one of the simplest and best-characterized protein systems. Samples consist of individual patches of two native purple-membrane monolayers (from here on referred as 2NPM), that is, 10 nm-thick stacks of two purple membranes, deposited on ultraflat gold surfaces. We focused on 2NPM, instead of monolayers, because we experimentally observed that prolonged contact-mode AFM measurements of 5 nm-thick membrane-monolayer patches yields to a degradation of our sample, which was evident from topography maps. IR difference-spectroscopy data are obtained with an AFM-based nanoscale IR technique that exploits the sample photothermal expansion due to IR absorption (ref [36], here abbreviated as AFM-IR). We achieve significant improvement in the signal-to-noise ratio (SNR) over existing IR nanospectroscopy by carefully choosing the power and the frequency tuning range of the IR laser, so as to obtain maximum photothermal expansion for each wavelength and by exploiting the field enhancement in the plasmonic nanogap that forms between the apex of a gold-coated AFM probe-tip and an ultraflat gold surface.[36] We demonstrate that when the nanogap is filled with 10 nm-thick 2NPM patches, AFM-IR can measure the IR difference-spectrum of BR embedded in native purple membranes, with a dramatic increase in the sensitivity per unit molecule compared with FTIR-based approaches. Analyzing our AFM-IR difference-spectra and comparing them to more conventional FTIR difference-spectra acquired on the same samples, we could identify two distinct photocycles that we attribute to TMPs that either adhere or not to the metal surfaces.

**Bacteriorhodopsin Photocycle.** BR absorbs photon energy via the retinal chromophore that is bound to a lysine of one of the helices via a protonated Schiff base, and acts as a proton pump moving protons from the cytoplasm to the extracellular space.[8] In Figure 1a we provide a simplified sketch of the BR photocycle. Photon absorption triggers a fast (∼100−200 fs) retinal isomerization from an all-trans to a 13-cis configuration, resulting in the intermediate state labeled J, which is converted in K and then in L.[37] After this sequence of fast transitions, the photocycle of wild-type BR can be described with the following important steps:[37] (i) proton transfer from the Schiff base to the amino acid Asp-85 acting as proton-acceptor (L to M transition); (ii) proton release toward the extracellular side in M (gray dashed arrow in Figure 1a); (iii) reprotonation of the Schiff base from the amino acid Asp-96 acting as proton-donor during M to N transition; (iv) reprotonation of Asp-96 from the cytoplasmatic surface in N (gray dashed arrow between the N and O intermediate states in Figure 1a); (v) retinal reisomerization (N to O transition), restoration of the initial photon absorption properties and deprotonation of Asp-85.

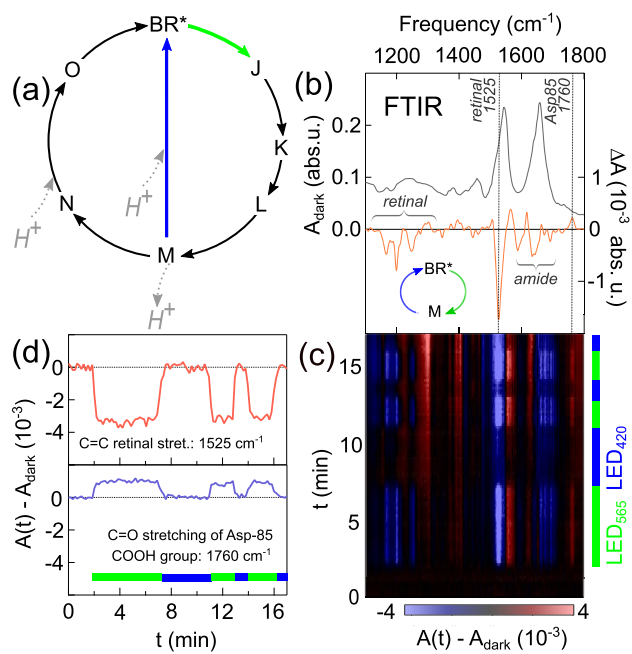

**Figure 1.** (a) Sketch of the BR photocycle. BR*: dark state; letters from J to O: intermediate states; $H^+$ and gray dashed arrow: proton release or capture. Green light initiates the photocycle (green arrow); in the D96N mutant, blue light brings proteins that accumulate in M back to BR* (blue arrow). (b) FTIR absorption spectrum of a thick film of native purple membranes in dark condition $A_{dark}$ (gray curve, left axis); corresponding difference-spectrum $\Delta A = A_{green} - A_{blue}$ (orange curve, right axis). (c) Time-dependent spectral plot of [$A(t) - A_{dark}$] acquired while alternately turning on and off the green and blue LEDs after a 2 min dark period, showing reproducible protein response. (d) Time-cuts taken from (c) at the negative peak of C=C retinal stretching (1525 cm$^{-1}$) and at the positive peak of C=O stretching of the COOH group of the proton acceptor Asp-85 (1760 cm$^{-1}$). Rise/decay times are a few seconds. A linear baseline has been subtracted from time-cuts in panel (d).

The capability of the AFM-IR technique to register the light-induced conformational changes of TMPs has been tested on the BR mutant D96N. In BR D96N, the replacement of the proton-donor Asp-96 with Asn-96, which cannot donate protons, prolongs the lifetime of the M intermediate state up to several seconds.[35] Since photoexcitation of M state with blue light leads back to a dark-state-like intermediate BR*,[8,38] one can then control the photocycle by alternating green/yellow and blue illuminations to induce the conversion BR* →

M and M → BR*. In our case we alternate illumination with green ($\lambda$ = 565 nm) and blue ($\lambda$ = 420 nm) light provided by loosely focused LED sources in order to control the photocycle of BR D96N and average the AFM-IR difference-spectra over many cycles to improve the SNR (see Methods).

**Fourier-Transform Infrared Difference-Spectroscopy.** In order to obtain a reference absorption spectrum $A_{dark}(\omega)$ of BR D96N membrane patches in the same form as those used for AFM-IR spectroscopy (poorly hydrated films on solid substrates), we performed FTIR transmission spectroscopy of thick films of purple membranes deposited by controlled solvent evaporation from membrane-patch suspensions on an IR-transparent CaF$_2$ window. In Figure 1b, the absorbance $A_{dark}(\omega)$ in dark condition (no illumination with visible light) is reported (gray curve, left axis). The spectrum clearly shows two amide bands associated with the helical structure of BR, that is, the amide-I band at ∼1660 cm$^{-1}$ and the amide-II band at ∼1540 cm$^{-1}$ predominantly assigned to the C=O stretching and N−H bending of the peptide bond, respectively.[9] Less intense bands associated with the lipid environment and with other amino-acid absorption peaks are seen at lower frequencies. The ratio between amide-I and amide-II intensities allows us to assume a partial stacking of the purple membranes, whose membrane plane is oriented mostly parallel to the surface of the CaF$_2$ substrate (see Supporting Information), although orientation of the cytoplasmatic/extracellular side cannot be assessed from IR data in general. The normal-incidence $\Delta A$ spectrum in Figure 1b then refers mostly to the contribution parallel to the membrane plane.

The FTIR difference-spectrum $\Delta A(\omega) = A_{green}(\omega) - A_{blue}(\omega)$ is obtained by consecutively illuminating the sample with green light and then with blue light, directly inside the FTIR spectrometer. In $\Delta A(\omega)$ (orange curve in Figure 1b, right axis), the pattern of negative and positive peaks between 1100 and 1800 cm$^{-1}$ is indicative of conformational changes of both the retinal and the protein backbone. In particular, the peak pattern between 1580 and 1700 cm$^{-1}$ is indicative of conformational changes of the protein backbone and the ethylenic modes of the retinal. The peaks between 1150 and 1300 cm$^{-1}$, instead, are assigned to the C−C retinal stretching, while the strong negative peak at 1525 cm$^{-1}$ and the less intense positive peak around 1560 cm$^{-1}$ are assigned to the C=C retinal stretching mode under blue and green light, respectively. The positive peak at 1760 cm$^{-1}$ in $\Delta A(\omega)$ of Figure 1b originates from the C=O stretching mode of the COOH functional group of the residue Asp-85, also an established spectroscopic marker of the proton transfer from the Schiff base to the proton acceptor Asp-85.[13,39,40] The spectral features in the $\Delta A(\omega)$ curve are clearly present also in the time-dependent difference spectral plots of Figure 1c,d, and they can be interpreted in terms of accumulation of a M-like intermediate state at green illumination,[39−41] even if in poor hydration conditions.[29,42,43]

**Infrared Difference-Spectra Acquired with AFM Tip-Enhanced Nanospectroscopy.** The AFM-IR platform, based on the coupling of an AFM and a tunable mid-IR quantum cascade laser (QCL), exploits the photothermal-induced mechanical resonance of the AFM cantilever, which, for soft samples and mid-IR laser excitation, has been definitively attributed to thermal expansion of the sample after IR radiation absorption.[36,44,45] The capability of AFM-IR to probe light-induced conformational changes of TMPs is first demonstrated with the setup in Figure 2a on the same purple membrane sample and visible illumination conditions as used for FTIR. A sample area of thickness $d \sim 1$ μm was located by AFM topography, and AFM-IR spectra were acquired under repeated 70 visible-light illumination cycles paying attention to keep the AFM tip located at the same sample position (see Methods). The gray curve in Figure 2b is the average of all AFM-IR spectra, and it represents the best possible estimate of the AFM-IR membrane absorption spectrum $A(\omega)$. AFM-IR difference spectra determined per each green/blue illumination cycle have been averaged to obtain $\Delta A$ shown in Figure 2b (purple curve, right axis). In Figure 2b, we also plot the $\Delta A$ spectrum for 70° incidence, that is, for an electric field orientation as in the AFM-IR experiment on thick samples, calculated by combining the normal incidence data in Figure 1b with oblique incidence FTIR data (see Supporting Information). One finds a satisfactory agreement in the entire frequency range, demonstrating that AFM-IR can indeed be used to probe the conformational changes of photosensitive proteins. In two frequency windows, the AFM-IR data are not reported due to the finite tuning range of the available QCL modules (see Methods).

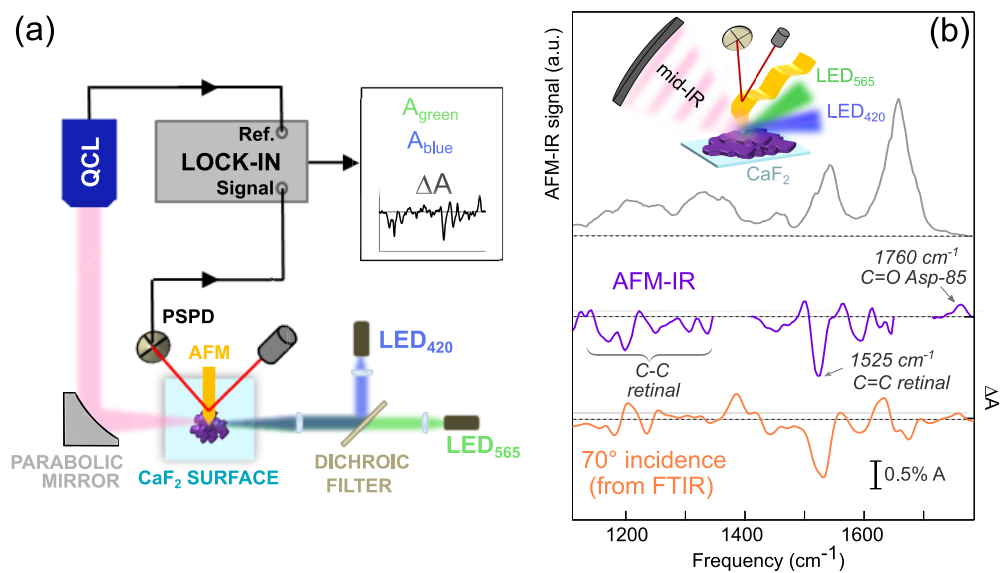

**Figure 2.** (a) Schematic of experimental setup for difference IR nanospectroscopy. QCL: quantum cascade laser, AFM: gold-coated atomic force microscope probe. (b) Gray curve: AFM-IR spectrum of a thick-film region ($d = 1$ μm) located in the purple-membrane assembly analyzed in Figure 1; violet curve: corresponding AFM-IR difference-spectrum $\Delta A$; orange curve: $\Delta A$ calculated for 70° incidence by combining the normal incidence data in Figure 1b with oblique incidence FTIR data.

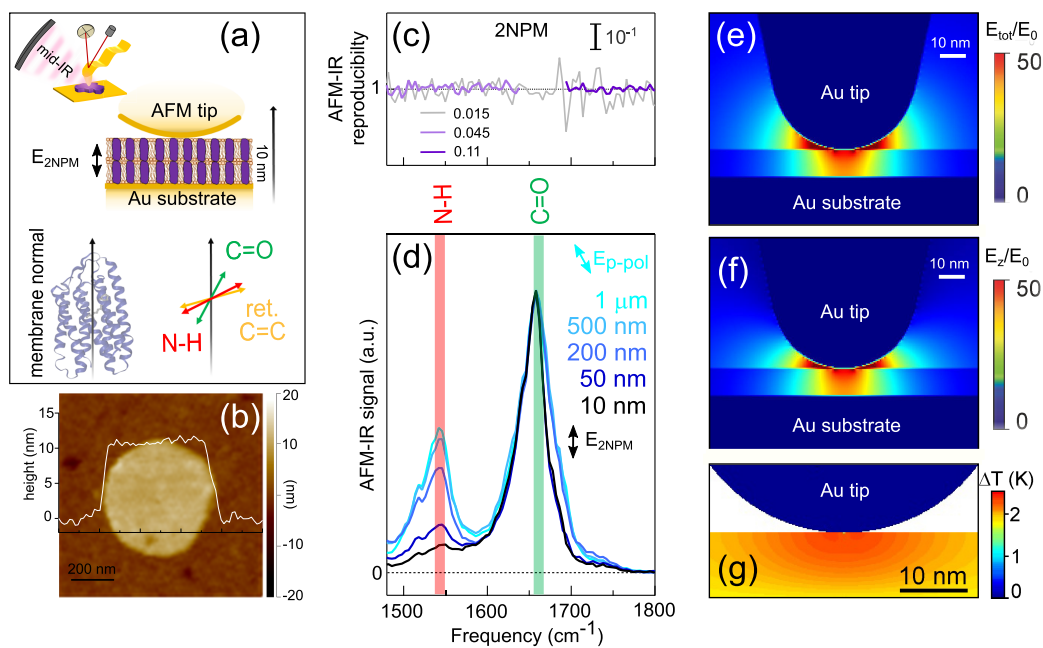

**Figure 3.** (a) Sketch of a 2NPM ($d = 10$ nm) located in the plasmonic nanogap between Au surface and Au-coated AFM tip; zoom of a BR molecule[67] (PDB ID:1FBB); color arrows indicate the direction of the main IR dipoles: C=O stretching, amide-I band;[51] N–H bending, amide-II band;[51] C=C stretching of retinal.[52] (b) AFM topography map of one 2NPM patch with superimposed topography profile. (c) Ratio between two subsequent AFM-IR spectra of a 2NPM in the full QCL tuning range (gray) and in two low-sample-absorption subranges (light and dark violet) changing the metal-mesh filters in front of the QCL so as to obtain maximum photothermal expansion in each range regardless the value of the sample absorption coefficient. (d) AFM-IR spectra acquired on purple-membrane films of different thickness $d$. An offset is subtracted so as to make each spectrum null at 1800 cm$^{-1}$ and the relative intensities are then normalized at 1660 cm$^{-1}$. E-field orientations are sketched for the p-polarized incident QCL beam ($E_{p\text{-pol}}$) and for the plasmonic nanogap ($E_{2NPM}$). (e–g) Simulated maps of: (e) E-field modulus; (f) E-field component normal to the membrane plane; and (g) temperature increase when the heat source is the absorbed radiation at 1540 cm$^{-1}$.

**Infrared Difference-Nanospectroscopy of a Single Purple-Membrane Patch.** AFM-IR difference-nanospectroscopy experiments were then performed on samples of decreasing thickness down to 2NPM patches, that is, 10 nm-thick stacks of two purple membranes. Note that, contrary to the techniques where an electrical signal is probed,[30] AFM-IR is not sensitive to the difference between cytoplasmatic and extracellular sides of the membrane. To obtain IR difference-spectra of isolated patches instead of thick films, the SNR of the AFM-IR technique needs to be increased by plasmonic field enhancement. To this aim, BR D96N purple membranes were drop-casted onto ultraflat template-stripped gold surfaces (Figure 3a). It has been demonstrated that, by depositing thin films on metallic surfaces, the gold-coated AFM tip provides a strong enhancement of the IR radiation intensity in the nanogap between the metallic tip apex and the surface,[36] together with a fully surface-normal orientation of the IR electric field below the tip apex. IR absorption spectra of molecular monolayers[36] and single purple membranes[19] have already been reported in the literature with the same resonantly enhanced AFM-IR technique[46] employed here, which makes use of QCLs pulsed at high repetition rate (~200 kHz), also limiting the transient sample heating to a few degrees.[36,47,46] A representative AFM topography map of a 2NPM patch is shown in Figure 3b. It reveals a thickness of $d = 10$ nm corresponding to two overlapping purple membranes and a flat top surface within the AFM sensitivity (RMS of 0.5 nm). In Figure 3d the AFM-IR spectra in the amide-I and amide-II regions are shown for samples of decreasing thickness and on different surfaces. One can appreciate that the SNR observed for 2NPMs is not significantly different from that observed for thick films, because the volume reduction of the probed material is compensated by the increased field intensity in the narrower plasmonic nanogap.[48,49] The electromagnetic and thermal simulations presented in Figure 3e–g quantitatively explain the origin of the AFM-IR photothermal expansion signal from a 10 nm-thick film with dielectric function equal to that of BR proteins. In Figure 3e,f, the field-enhancement for the total electric field $E_{tot}$ and for the surface-normal component $E_z$ are compared using the same color scale, and one can appreciate that a negligible difference exists in the nanoscale sample volume precisely located under the apex tip (approximately, a cylinder of 20 nm radius in which the field is oriented normal to the surface). This is also the volume of highest photoinduced temperature increase $\Delta T$ (see Figure 3g). Simulations were performed with IR radiation at 1540 cm$^{-1}$, but no major qualitative differences are expected over the entire 1500–1800 cm$^{-1}$ range, apart from the value of $\Delta T$ which depends on both the values of the frequency-dependent absorption coefficient and of the emitted laser power. For focused laser power of 10 mW, one obtains from simulations $\Delta T \sim 2$ K at 1540 cm$^{-1}$ at the end of each 260 ns-long QCL pulse. The plot in Figure 3g suggests that, in this specific thin-film-on-metal-surface configuration and at wavelengths where the sample absorption coefficient is high, the sample volume expected to generate the AFM-IR signal corresponds to the cylinder of 20 nm radius and height 10 nm, containing approximately 150 BR trimers, which represent the unit cell of the native assembling of BR molecules in purple membranes.[50]

The role of dipole selection rules is evident in Figure 3d. The dipole of the amide-II and of the C=C retinal stretching are mostly parallel to the membrane plane (red and yellow arrows in Figure 3a indicating ~70° and ~73° with the surface normal for N–H bending[51] and C=C stretching,[52] respectively) and therefore the signal intensity is suppressed in the 1500–1600 cm$^{-1}$ range with decreasing thickness (red shaded area in Figure 3d) due to increasingly surface-normal direction of the radiation electric field. Vice versa, the amide-I dipole is mostly oriented normal to the membrane plane[51] (~40° with the surface normal, green arrow in Figure 3a), so the intensity of the 1660 cm$^{-1}$ peak related to α helix increases by 15% in the 2NPM (surface-normal electric field) if compared to the p-polarized incidence on thick films, where the electric field is oriented at 20° with the surface normal (note that in Figure 3d, all spectra have been normalized to the amide-I peak). The modifications of the amide-I line shape with varying electric field orientation seen in Figure 3d are in agreement with previous IR linear dichroism studies of BR[49,52] and are discussed in the Supporting Information.

In the C=O stretching region around 1760 cm$^{-1}$, the absolute value of the AFM-IR signal is very low, but the 45° orientation of the C=O bond of Asp-85[53] makes it anyway possible to observe relevant difference signals from 2NPMs. Normally the laser power in AFM-IR is set to a low value to avoid saturation at the frequency of peak absorption (here 1660 cm$^{-1}$). In order to further boost the SNR beyond the use of plasmonic field enhancement,[36] we optimized the spectral acquisition procedure by adjusting the IR laser power, up to 50 mW peak power, according to the value of the absorption and to the frequency-dependent laser emission power in each range. The result of this procedure is evident in Figure 3c where we compare the AFM-IR reproducibility line obtained for two spectra acquired in the full 1480–1800 cm$^{-1}$ range with a metal-mesh filter with transmittance of 0.015, with that obtained in the frequency intervals 1480–1640 cm$^{-1}$ and 1690–1800 cm$^{-1}$, where the sample absorption is low, with a second and third metal-mesh filters with transmittance of 0.045 and 0.11, respectively. The standard deviation of the reproducibility lines passes from ~3% at low power, to ~1% for high power, enabling us to reach a SNR of ~$10^3$ by averaging over 70 difference-spectra acquisitions (see Methods).

**Discussion of Difference-Spectra.** In Figure 4a,b the AFM-IR difference-spectra $\Delta A$ are shown for different film thickness $d$, from 1 μm down to the 2NPM ($d = 10$ nm). Apart from the $d = 1$ μm spectrum taken from Figure 2b and measured on a CaF$_2$ window, all other spectra were taken on ultraflat gold surfaces. Since AFM-IR technique does not provide an absolute value of the absorbance, the $\Delta A$ spectra in Figure 4 are normalized using an arbitrary normalization coefficient so as to equal the difference spectra at 1590 cm$^{-1}$, a wavenumber at which no dichroic behavior is usually observed.[53,54] Turning our attention to the AFM-IR spectra for the 2NPM ($d = 10$ nm) in Figure 4a,b, we clearly observe three main variations arising in $\Delta A$: (1) the intensity of the negative peak of C=C of retinal at ~1525 cm$^{-1}$ is reduced in the 2NPM down to 50% compared to the thick film ($d = 1$ $\mu$m) intensity; (2) a new negative peak appears at ∼1545 cm$^{-1}$; (3) the intensity of the positive peak of C=O of Asp-85 at 1760 cm$^{-1}$ is reduced in the 2NPM down to 50% of the $d = 1$ $\mu$m intensity (see Figure 4a,b where the same numbering has been used to highlight the three variations). The spectrum for $d = 50$ nm presents the same features as the 2NPM spectrum, but less intense, suggesting that the observed changes are indeed related to the film thickness and possibly to the nature of the substrate. The $d = 1$ $\mu$m and $d = 10$ nm $\Delta A$ are expected to slightly differ in terms of relative intensity of positive and negative peaks due to the selection rule effect. Variations between the difference-spectra acquired under different anisotropic polarization conditions can also be due to dipole tilts during the BR photocycle. In the Supporting Information, these effects are discussed in detail: in particular, the effect of the anisotropic field orientation is discussed by comparing micro-FTIR difference spectra and calculations performed with a hybrid quantum/classical approach.[55] The variations (1), (2), and (3) observed between the $d = 1$ $\mu$m and the $d = 10$ nm sample in Figure 4 are simply too large to be interpreted as a selection rule effect or as a dipole tilt, or as a combination of the two. Note that the electric field orientation changes only by 20° degrees between the $d = 1$ $\mu$m and the $d = 10$ nm sample, and that both the C=C bond of retinal (1525 cm$^{-1}$ negative peak) and the C=O bond of Asp-85 (1760 cm$^{-1}$ peak) do not significantly change their orientation during the photocycle.[53,54]

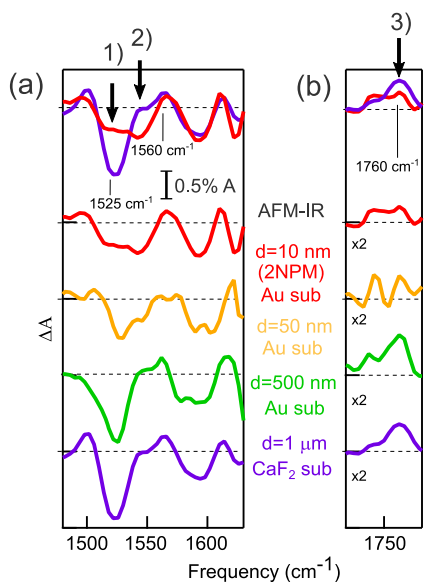

The suppression of the Asp-85 proton acceptor peak at 1760 cm$^{-1}$ seen in the case of 2NPMs, that is, variation (3) in Figure 4, may indicate that light-activated proton transport is partly prevented in proteins embedded in the 2NPM.[17] This interpretation would also be in agreement with the reduced photoelectrical activity often observed for BR molecules adhering to solid surfaces by AFM-based and electrochemistry techniques.[30,56,57] In particular, a clear distinction between the photoelectrical activity of BR monolayers and multilayers has been verified in ref 30, pointing to an alteration of BR photocycle in the case of monolayers adhering on a solid surface compared to multilayers and suspensions. The impossibility for net proton transport through the membrane within a photocycle is typically accompanied by modifications of the retinal and protein conformational changes if compared to those of the standard BR photocycle.[8,58] We therefore suggest that the variations (1) and (2) in Figure 4 can be interpreted as the spectral markers of a modified conformation of the retinal and/or retinal Schiff base and/or its environment, occurring under blue light illumination. In the standard photocycle of BR, the retinal isomerization induced by photon absorption results in the shift of the C═C vibrational frequency from its value around 1525 cm$^{-1}$ in the dark state BR* to a new value around 1560 cm$^{-1}$ for the M intermediate state[59] (see Figure 4a). In the 2NPM spectrum, we instead observe a branching of the negative peak at ∼1525 cm$^{-1}$ into a second negative peak at ∼1545 cm$^{-1}$, that is, variation (2). This feature can be assigned to a modified frequency for the retinal C═C absorption in a fraction of proteins that under blue illumination are not brought back to the dark state BR* (C═C absorption at ∼1525 cm$^{-1}$), but rather, they correspond to a nonstandard intermediate state characterized by a C═C absorption at ∼1545 cm$^{-1}$. One can then deduce a branching of the photocycle: only a fraction of proteins follows the standard photocycle contributing to the net proton transport through the membrane, while the remaining proteins undergo a different photocycle branch with zero net proton transport through the membrane, which is characterized by retinal C═C vibration frequency at ∼1560 cm$^{-1}$ and ∼1545 cm$^{-1}$ under green and blue light illumination, respectively. The intensity of the Asp-85 peak at 1760 cm$^{-1}$, which is proportional to the net proton transport through the membrane, is found to be suppressed to 50% in the 2NPM with respect to its thick-film value, therefore we can estimate that 50% of proteins participate to the standard photocycle and 50% to the photocycle branch.

**Figure 4.** (a,b) AFM-IR difference-spectra of purple-membrane films of thickness $d$ deposited either on CaF$_2$ or Au. Of the full QCL tuning range 1480−1800 cm$^{-1}$, two subranges with high SNR are shown in (a) and (b) respectively. The worse SNR for $d = 50$ nm is due to poor tip indentation in that specific measurement. On top of panels (a) and (b) the $\Delta A$ for the 2NPM ($d = 10$ nm, red curve) and for the thick film ($d = 1$ $\mu$m, violet curve) are superimposed to highlight the changes arising in the 2NPM spectrum at 1525, 1545 and 1760 cm$^{-1}$, respectively labeled by the numbers 1, 2, and 3 used in the text to identify them.

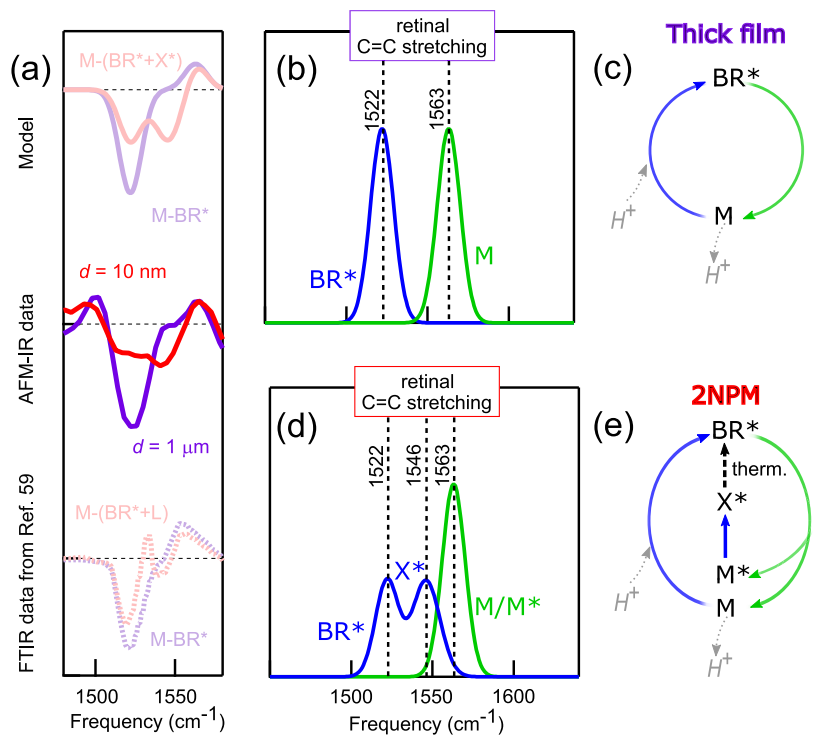

**Figure 5.** (a) Comparison of the AFM-IR $\Delta A$ data for $d = 1$ $\mu$m (thick film) and $d = 10$ nm (2NPM patch) with the calculated $\Delta A$ from a Gaussian line shape model of the C=C absorption of retinal. The model absorption spectra are reported in panels (b) and (d) for thick film and 2NPM, respectively. The simplified model photocycles for thick films and 2NPMs, as derived from the line shape analysis, are shown in panels (c) and (e), respectively. In the bottom part of panel (a), two relevant difference-spectra calculated from literature FTIR data[59] (see text) are also shown for comparison (dotted curves).

In Figure 5, we present a model of the difference-spectroscopy line shape of the C═C retinal through the branching of the photocycle for 2NPMs. For thick films, the C═C retinal absorption band is modeled as a Gaussian line shape rigidly shifting from 1522 cm$^{-1}$ under blue light to 1563 cm$^{-1}$ under green light in the standard BR photocycle (Figure 5b). For 2NPMs under blue light illumination, the C═C retinal absorption band splits into two sub-bands of equal intensity centered at 1522 and 1546 cm$^{-1}$ (Figure 5d). The peak at 1546 cm$^{-1}$ represents the retinal C═C absorption of the protein fraction that are not brought back to the dark state BR* but are instead trapped in a different intermediate state that we label X*. The simplified model photocycles for the thick film and the 2NPM are sketched in Figure 5c,e. For 2NPMs, the M intermediate population is split into two subpopulations that we label M and M*. The C═C retinal absorption band of M and M* is identical, but upon absorption of blue light, M follows the standard proton-pump photocycle, while M* is transformed into the X* intermediate state. Our photocycle branching model also assumes that proteins falling in the X* intermediate state thermally decay into BR*, insofar reproducing the same final state as the standard BR D96N photocycle under green/blue illumination cycles. In the bottom part of Figure 5a, we also present a comparison of our data and our model with the difference-spectra calculated from the extensive FTIR data of ref 59 characterizing the IR spectra of all BR intermediate states. Here we employ the spectrum of the L intermediate as a model for the X* spectrum due to the similarity of the C═C absorption frequency with the X* intermediate state,[59] and we calculate the difference-spectrum of the mixture of BR* and L states (dotted pink curve in Figure 5a), to be compared with the experimental spectrum of the mixture of BR* and X* states (red curve in Figure 5a). Summarizing, IR nanospectroscopy has enabled us to characterize at the nanoscale a protein conformational change, related to the inhibition of the proton pump activity at the membrane monolayer level when in contact with metal surfaces. The alteration of the photocycle for proteins located in the tip−surface nanogap may be ascribed to several mechanisms, such as the inhibition of the standard protein conformational changes and reprotonation process due to less flexibility of those proteins directly adhering to the surface, or alteration of membrane surface potential due to the presence of the gold surface and the metallic AFM tip. However, extended experimental validations would be necessary in order to achieve a complete interpretation of the results, and this would be beyond the scope of this paper.

In conclusion, we have demonstrated that atomic-force-microscope-assisted infrared difference-spectroscopy with plasmonic field enhancement at the probe tip can be fruitfully employed to identify subtle conformational changes of transmembrane proteins in nanoscale samples. The bacteriorhodopsin photocycle is studied in 2NPM cell membrane patches, whose dimensions are well below the diffraction limit for conventional infrared difference-spectroscopy. The number of protein molecules probed in the mid-infrared, of the order of ∼500, is comparable to that reached by spectroscopies based on visible light lasers such as Förster resonance energy transfer and micro-Raman spectroscopy, also capable of measuring conformational changes. However, in the case of light-sensitive proteins, those methods, which are based on visible-light lasers, cannot be directly employed to study the native membrane photocycle. Here instead we could observe a branching of the photocycle, accounting for less than 1% relative infrared absorption change at the retinal stretching mode frequency of 1525 cm$^{-1}$, when the bacteriorhopsin molecules are in close contact to metal surfaces. The effect could be attributed either to mechanical inhibition of the conformational change or to electrostatic potential arising from physisorption. The obtained results are relevant for bioelectronic and biophotonic devices, in which rhodopsin molecules are put in contact with metal surfaces (e.g., electrodes).

In perspective, the presence of an atomic force microscopy probe in emerging nanoinfrared techniques could be exploited well beyond its plasmonic field-enhancement capabilities. For example, the metal-coated tip could be used to introduce an electric potential with respect to the solid surface, so as to provide electrophysiology techniques with nanoscale resolution and simultaneous spectroscopic information on transmembrane protein conformational changes. Also, the future implementation of liquid environments[34,60] and/or controlled atmospheres in nanoinfrared techniques could enable the nanospectroscopy study of relevant biological processes involving transmembrane gate proteins, such as optogenetics and photoinduced drug delivery.

## ■ METHODS

**Sample Preparation.** Patches of purple membranes densely filled with BR D96N were purified as reported in ref 61 and stored in buffer solution (20 × 10$^{-3}$ M Bis-Tris propane, 100 × 10$^{-3}$ M NaCl, 1 × 10$^{-3}$ M MgCl$_2$) at −80 °C. Droplets of the suspension containing purple membranes were either cast onto IR-transparent CaF$_2$ windows or onto 1 cm × 1 cm template-stripped gold chips (Platypus Technologies, 0.3 nm rms roughness) and let dry in air. After 10 min, the samples deposited on ultraflat gold surfaces were rinsed with the buffer solution (phosphate buffered saline at pH = 7.8) and subsequently dried for 1 h in an atmosphere with humidity below 10%.

**Visible Illumination.** Light to control the BR photocycle was provided by two LEDs (Thorlabs M565L3, center wavelength at 565 nm (green) and M420L3, 420 nm (blue)). The LED output beams were made collinear using lenses and a dichroic filter (Thorlabs - MD480) and then focused on the sample. The power density for both LEDs was ∼50 mW/cm$^2$. Sample was light-adapted before all experiments, and therefore, the photocycle starts from homogeneous isomeric dark state (all-trans).

**FTIR Spectroscopy.** FTIR measurements were performed in transmission mode with either a Bruker Vertex 70v (Figure 1b) or a Bruker Vertex 80v (Figure 1c) equipped with blackbody sources and HgCdTe detectors and in vacuum condition. The transmitted intensity was collected both in dark condition ($T_{dark}$) or under visible illumination ($T_{blue}$ or $T_{green}$) by averaging 512 interferometer scans at 8 cm$^{-1}$ spectral resolution. The $A_{dark}$ curve of Figure 1b was then calculated as $A_{dark} = -\log(T_{dark}/T_0)$ using pristine CaF$_2$ as the reference channel ($T_0$). The $\Delta A$ curve of Figure 1b was calculated as $\Delta A = A_{green} - A_{blue} = \log(T_{blue}/T_{green})$, as there is no dependence of the reference $T_0$ on the visible-light illumination. The $A(t) - A_{dark}$ spectra of Figure 1c were calculated as $A(t) - A_{dark} = \log(T_{dark}/T(t))$ where $T_{dark}$ is the transmission spectrum obtained by averaging over the spectra acquired during the first 2 min in dark condition.

Oblique incidence micro-FTIR spectra were obtained using a gas-purged Nicolet Continuum Infrared Microscope with a HgCdTe detector, a 15× Cassegrain reflective objective with spherical aberration compensation (numerical aperture NA = 0.58) and a knife-edge aperture of 100 μm × 100 μm. Reflection of the gold surface covered with the membrane films in dark condition $R_{dark}$ was acquired first, then $R_{blue}$ and $R_{green}$ were acquired. A total of 1024 interferometer scans at 8 cm$^{-1}$ spectral resolution were accumulated for each spectrum. Reflection from the pristine gold surface $R_0$ was used as reference to calculate the double-transmission spectrum of the membrane film $R_{dark}/R_0$ and the absorption $A_{dark} = -\log(R_{dark}/R_0)$. The micro-FTIR difference-spectra $\Delta A$ were calculated as $\Delta A = A_{green} - A_{blue} = \log(R_{blue}/R_{green})$.

**AFM-IR Setup.** AFM-IR measurements were performed using a platform of tip-enhanced IR nanospectroscopy based on the photothermal expansion effect (NanoIR2, Anasys Instruments). AFM-IR measurements were performed by purging with dry air both the optics and the sample compartment for many hours. The p-polarized beam from a broadly tunable mid-IR QCL (MIRCATxB, Daylight Solutions, with a spectral range 900–1800 cm$^{-1}$) is tightly focused onto the gold-coated tip of an AFM probe. The IR beam impinges from the side at an angle of 70° to the surface normal, that is, with the electric filed oriented at 20° with surface normal. The laser provided 260 ns long light pulses at a repetition frequency that was chosen as to be in resonance with the second mechanical bending mode of the cantilever at ∼200 kHz (resonantly enhanced infrared nanospectroscopy, REINS[36]). The laser power was adjusted using transmission metal-mesh filters in front of the QCL output and it was in the range 5−50 mW depending on frequency range and sample absorption.

**AFM-IR Difference-Spectroscopy: Spectral Acquisition Protocol.** $\Delta A$ were obtained with AFM-IR by averaging over 70 difference-spectra $A_{green}(\omega) - A_{blue}(\omega)$. The AFM-IR spectra were acquired with 4 cm$^{-1}$ steps, ∼ 0.3 s per step subdivided in integration time and tuning time. The $\Delta A$ spectra of Figure 2b and Figure 4 were obtained by stitching the data acquired in different subranges where the laser power was tuned in order to have comparable AFM-IR signal intensity regardless of the absolute value of the sample absorption that depends on electric field orientation. After each 5 green/blue acquisitions, a topography map of the sample area was acquired, in order to place the AFM tip exactly at the same position for all acquisitions.

The AFM-IR $\Delta A$ could not be acquired in two narrow frequency ranges around 1430 and 1680 cm$^{-1}$ corresponding to the boundaries of the tuning ranges of the individual laser chips in the MIRcat system, in which fluctuations of the laser emission power are highest.

**Electromagnetic and Thermal Simulations.** The finite-difference time-domain method (FDTD Solutions, Lumerical Inc., Canada) has been used to simulate a sample consisting of a 10 nm-thick dielectric film (representing the 2NPM) on top of a 100 nm-thick Au layer, on top of 2 μm-thick epoxy layer. The AFM tip was modeled as a smoothed-tip cone with half apical angle of 12° and apex curvature radius of 25 nm, following scanning electron microscope images of the probe. The tip is made of Si with a 25 nm-thick Au coating. The IR beam is modeled as a weakly focused Gaussian beam incident at 70° to the surface normal with peak power density of 8 × 10$^6$ W/m$^2$. The 10 nm-thick film is assumed to have complex refractive index $n = 1.7 + 0.45i$, taken at the frequency of the amide-II band (∼1540 cm$^{-1}$) from the out-of-plane component of the refractive index (i.e., orthogonal to the membrane plane) reported in ref 62. Thermal simulations are performed with a commercial heat transport solver (HEAT, Lumerical Inc., Canada) by imposing a fixed temperature (room temperature) boundary condition at both the lower surface of the substrate and the upper surface of the AFM tip. For the 2NPM, we used a thermal conductivity of 0.25 W m$^{-1}$K$^{-1}$ (ref 63), a mass density of 1250 kg m$^{-3}$ (calculated assuming a protein mass density of 1400 kg m$^{-3}$ (ref 64), a lipid mass density of 800 kg m$^{-3}$ (ref 65), a protein:lipid ratio of 75:25), and a heat capacity of 2.4 × 10$^3$ J kg$^{-1}$ K$^{-1}$ (ref 66). In order to qualitatively mimic the unavoidable presence of a thermal interface resistance between the tip and the membrane, we introduce a thin (0.1 nm) film with a low thermal conductivity of about 0.0025 W m$^{-1}$ K$^{-1}$. The extension of the nanoscale colder spot located below the tip apex (an effect of the thermal reservoir represented by the tip) depends on the poorly characterized properties of such a surface resistance, while the maximum photoinduced temper-ature increase reached in the membrane changes by less than 20% after the introduction of the surface insulating layer.

**IR Data Analysis.** A smoothing spline algorithm and a baseline correction have been applied to obtain the final AFM-IR $\Delta A$ curves reported in Figure 2 and Figure 4 (see Supporting Infromation).

## ASSOCIATED CONTENT

### Supporting Information

The Supporting Information is available free of charge on the ACS Publications website at DOI: 10.1021/acs.nanolett.9b00512.

> Native purple-membrane orientation in our samples; effect of dipole selection rules on the difference-spectra; comparison between micro-FTIR data and spectra calculation in the amide-I band frequency interval; model for photocycle branching; smoothing and baseline correction of AFM-IR data; O−H stretching region (PDF)

## AUTHOR INFORMATION

**Corresponding Authors**

*E-mail: michele.ortolani@roma1.infn.it.
*E-mail: valeria.giliberti@iit.it.

**ORCID**

Peter Hegemann: 0000-0003-3589-6452
Laura Zanetti-Polzi: 0000-0002-2550-4796
Isabella Daidone: 0000-0001-8970-8408
Stefano Corni: 0000-0001-6707-108X
Paolo Biagioni: 0000-0003-4272-7040
Leonetta Baldassarre: 0000-0003-2217-0564
Michele Ortolani: 0000-0002-7203-5355

**Author Contributions**

M.B. and P.H. prepared the purple membrane samples. V.G. and R.P. constructed the illumination setup and performed the AFM-IR experiments. V.G., M.O., E.R., L.P., and U.S. performed the FTIR experiments. I.D., L.Z.-P., and S.C. calculated the protein vibrational spectra. F.R. and P.B. performed electromagnetic and thermal simulations. M.O. and L.B. designed the experiment and coordinated the work. V.G. and M.O. wrote the manuscript. All authors contributed extensively to the analysis and discussion of the results.

**Funding**

M.O. and L.B. acknowledge funding by Sapienza University of Rome through the program Ricerca d'Ateneo 2017 (Grant No. PH11715C7E435F41). S.C. acknowledges funding by the European Research Council, under the grant ERC-CoG-681285 (TAME-Plasmons). The research leading to this result has been supported by the project CALIPSOplus under the Grant Agreement 730872 from the EU Framework Programme for Research and Innovation HORIZON 2020. E.R. and P.H. acknowledge support from the German Federal Ministry of Education and Research (BMBF) Grant No. 05K16KH1. M.B. and P.H. were supported by the German Research Society (DFG, SFB1078) and P.H. was supported by the Hertie foundation.

**Notes**

The authors declare no competing financial interest.

## ACKNOWLEDGMENTS

Measurements were also carried out at the IRIS beamline at Helmholtz-Zentrum Berlin für Materialien und Energie.

## REFERENCES

(1) Ernst, O. P.; Lodowski, D. T.; Elstner, M.; Hegemann, P.; Brown, L. S.; Kandori, H. Microbial and Animal Rhodopsins: Structures, Functions, and Molecular Mechanisms. *Chem. Rev.* **2014**, *114* (1), 126−163.

(2) Wald, G. The molecular basis of visual excitation. *Nature* **1968**, *219* (5156), 800.

(3) Scheib, U.; Broser, M.; Constantin, O. M.; Yang, S.; Gao, S.; Mukherjee, S.; Stehfest, K.; Nagel, G.; Gee, C. E.; Hegemann, P. Rhodopsin-cyclases for photocontrol of cGMP/cAMP and 2.3 Å structure of the adenylyl cyclase domain. *Nat. Commun.* **2018**, *9* (1), 2046.

(4) Oesterhelt, D.; Meentzen, M.; Schuhmann, L. Reversible Dissociation of the Purple Complex in Bacteriorhodopsin and Identification of 13-Cis and All-Trans-Retinal as Its Chromophores. *Eur. J. Biochem.* **1973**, *40* (2), 453−463.

(5) Jiang, Y.; Lee, A.; Chen, J.; Cadene, M.; Chait, B. T.; MacKinnon, R. The Open Pore Conformation of Potassium Channels. *Nature* **2002**, *417* (6888), 523−526.

(6) Nagel, G.; Szellas, T.; Huhn, W.; Kateriya, S.; Adeishvili, N.; Berthold, P.; Ollig, D.; Hegemann, P.; Bamberg, E. Channelrhodopsin-2, a Directly Light-Gated Cation-Selective Membrane Channel. *Proc. Natl. Acad. Sci. U. S. A.* **2003**, *100* (24), 13940−13945.

(7) Lozier, R. H.; Bogomolni, R. A.; Stoeckenius, W. Bacteriorhodopsin: A Light-Driven Proton Pump in Halobacterium Halobium. *Biophys. J.* **1975**, *15* (9), 955−962.

(8) Haupts, U.; Tittor, J.; Oesterhelt, D. Closing in on Bacteriorhodopsin: Progress in Understanding the Molecule. *Annu. Rev. Biophys. Biomol. Struct.* **1999**, *28* (1), 367−399.

(9) Barth, A. Infrared Spectroscopy of Proteins. *Biochim. Biophys. Acta, Bioenerg.* **2007**, *1767* (9), 1073−1101.

(10) Ritter, E.; Puskar, L.; Bartl, F. J.; Aziz, E. F.; Hegemann, P.; Schade, U. Time-Resolved Infrared Spectroscopic Techniques as Applied to Channelrhodopsin. *Front. Mol. Biosci.* **2015**, *2*, 38.

(11) Alexiev, U.; Farrens, D. L. Fluorescence Spectroscopy of Rhodopsins: Insights and Approaches. *Biochim. Biophys. Acta, Bioenerg.* **2014**, *1837* (5), 694−709.

(12) Kottke, T.; Lórenz-Fonfría, V. A.; Heberle, J. The Grateful Infrared: Sequential Protein Structural Changes Resolved by Infrared Difference Spectroscopy. *J. Phys. Chem. B* **2017**, *121* (2), 335−350.

(13) Lórenz-Fonfría, V. A.; Saita, M.; Lazarova, T.; Schlesinger, R.; Heberle, J. PH-Sensitive Vibrational Probe Reveals a Cytoplasmic Protonated Cluster in Bacteriorhodopsin. *Proc. Natl. Acad. Sci. U. S. A.* **2017**, *114* (51), E10909−E10918.

(14) Lórenz-Fonfría, V. A.; Schultz, B.-J.; Resler, T.; Schlesinger, R.; Bamann, C.; Bamberg, E.; Heberle, J. Pre-Gating Conformational Changes in the ChETA Variant of Channelrhodopsin-2 Monitored by Nanosecond IR Spectroscopy. *J. Am. Chem. Soc.* **2015**, *137* (5), 1850−1861.

(15) Nyquist, R. M.; Ataka, K.; Heberle, J. The Molecular Mechanism of Membrane Proteins Probed by Evanescent Infrared Waves. *ChemBioChem* **2004**, *5* (4), 431−436.

(16) Neubrech, F.; Huck, C.; Weber, K.; Pucci, A.; Giessen, H. Surface-Enhanced Infrared Spectroscopy Using Resonant Nanoantennas. *Chem. Rev.* **2017**, *117* (7), 5110−5145.

(17) Jiang, X.; Zaitseva, E.; Schmidt, M.; Siebert, F.; Engelhard, M.; Schlesinger, R.; Ataka, K.; Vogel, R.; Heberle, J. Resolving Voltage-Dependent Structural Changes of a Membrane Photoreceptor by Surface-Enhanced IR Difference Spectroscopy. *Proc. Natl. Acad. Sci. U. S. A.* **2008**, *105* (34), 12113−12117.

(18) Ataka, K.; Stripp, S. T.; Heberle, J. Surface-Enhanced Infrared Absorption Spectroscopy (SEIRAS) to Probe Monolayers of Membrane Proteins. *Biochim. Biophys. Acta, Biomembr.* **2013**, *1828* (10), 2283−2293.

(19) Giliberti, V.; Badioli, M.; Nucara, A.; Calvani, P.; Ritter, E.; Puskar, L.; Aziz, E. F.; Hegemann, P.; Schade, U.; Ortolani, M.; Baldassarre, L. Heterogeneity of the Transmembrane Protein Conformation in Purple Membranes Identified by Infrared Nanospectroscopy. *Small* **2017**, *13* (44), 1701181.

(20) Carpenter, E. P.; Beis, K.; Cameron, A. D.; Iwata, S. Overcoming the Challenges of Membrane Protein Crystallography. *Curr. Opin. Struct. Biol.* **2008**, *18* (5), 581−586.

(21) Pfreundschuh, M.; Harder, D.; Ucurum, Z.; Fotiadis, D.; Müller, D. J. Detecting Ligand-Binding Events and Free Energy Landscape While Imaging Membrane Receptors at Subnanometer Resolution. *Nano Lett.* **2017**, *17* (5), 3261−3269.

(22) Pfitzner, E.; Seki, H.; Schlesinger, R.; Ataka, K.; Heberle, J. Disc Antenna Enhanced Infrared Spectroscopy: From Self-Assembled Monolayers to Membrane Proteins. *ACS Sensors* **2018**, *3* (5), 984−991.

(23) Dufrêne, Y. F.; Ando, T.; Garcia, R.; Alsteens, D.; Martinez-Martin, D.; Engel, A.; Gerber, C.; Müller, D. J. Imaging modes of atomic force microscopy for application in molecular and cell biology. *Nat. Nanotechnol.* **2017**, *12* (4), 295.

(24) Shibata, M.; Yamashita, H.; Uchihashi, T.; Kandori, H.; Ando, T. High-Speed Atomic Force Microscopy Shows Dynamic Molecular Processes in Photoactivated Bacteriorhodopsin. *Nat. Nanotechnol.* **2010**, *5* (3), 208−212.

(25) Casuso, I.; Fumagalli, L.; Samitier, J.; Padrós, E.; Reggiani, L.; Akimov, V.; Gomila, G. Nanoscale Electrical Conductivity of the Purple Membrane Monolayer. *Phys. Rev. E* **2007**, *76* (4), 041919.

(26) Berthoumieu, O.; Patil, A. V.; Xi, W.; Aslimovska, L.; Davis, J. J.; Watts, A. Molecular Scale Conductance Photoswitching in Engineered Bacteriorhodopsin. *Nano Lett.* **2012**, *12* (2), 899−903.

(27) Muller, D. J. AFM: A Nanotool in Membrane Biology. *Biochemistry* **2008**, *47* (31), 7986−7998.

(28) Renugopalakrishnan, V.; Barbiellini, B.; King, C.; Molinari, M.; Mochalov, K.; Sukhanova, A.; Nabiev, I.; Fojan, P.; Tuller, H. L.; Chin, M.; Somasundaran, P.; Padrós, E.; Ramakrishna, S. Engineering a Robust Photovoltaic Device with Quantum Dots and Bacteriorhodopsin. *J. Phys. Chem. C* **2014**, *118* (30), 16710−16717.

(29) Mohammadpour, R.; Janfaza, S. Efficient Nanostructured Biophotovoltaic Cell Based on Bacteriorhodopsin as Biophotosensitizer. *ACS Sustainable Chem. Eng.* **2015**, *3* (5), 809−813.

(30) He, T.; Friedman, N.; Cahen, D.; Sheves, M. Bacteriorhodopsin Monolayers for Optoelectronics: Orientation and Photoelectric Response on Solid Supports. *Adv. Mater.* **2005**, *17* (8), 1023−1027.

(31) Berweger, S.; Nguyen, D. M.; Muller, E. A.; Bechtel, H. A.; Perkins, T. T.; Raschke, M. B. Nano-Chemical Infrared Imaging of Membrane Proteins in Lipid Bilayers. *J. Am. Chem. Soc.* **2013**, *135* (49), 18292−18295.

(32) Amenabar, I.; Poly, S.; Nuansing, W.; Hubrich, E. H.; Govyadinov, A. A.; Huth, F.; Krutokhvostov, R.; Zhang, L.; Knez, M.; Heberle, J.; Bittner, A. M.; Hillenbrand, R. Structural analysis and mapping of individual protein complexes by infrared nanospectroscopy. *Nat. Commun.* **2013**, *4*, 2890.

(33) Ruggeri, F. S.; Longo, G.; Faggiano, S.; Lipiec, E.; Pastore, A.; Dietler, G. Infrared nanospectroscopy characterization of oligomeric and fibrillar aggregates during amyloid formation. *Nat. Commun.* **2015**, *6*, 7831.

(34) Ramer, G.; Ruggeri, F. S.; Levin, A.; Knowles, T. P.; Centrone, A. Determination of Polypeptide Conformation with Nanoscale Resolution in Water. *ACS Nano* **2018**, *12* (7), 6612−6619.

(35) Holz, M.; Drachev, L. A.; Mogi, T.; Otto, H.; Kaulen, A. D.; Heyn, M. P.; Skulachev, V. P.; Khorana, H. G. Replacement of Aspartic Acid-96 by Asparagine in Bacteriorhodopsin Slows Both the Decay of the M Intermediate and the Associated Proton Movement. *Proc. Natl. Acad. Sci. U. S. A.* **1989**, *86* (7), 2167−2171.

(36) Lu, F.; Jin, M.; Belkin, M. A. Tip-Enhanced Infrared Nanospectroscopy via Molecular Expansion Force Detection. *Nat. Photonics* **2014**, *8* (4), 307−312.

(37) Lanyi, J. K. Proton transfers in the bacteriorhodopsin photocycle. *Biochim. Biophys. Acta, Bioenerg.* **2006**, *1757* (8), 1012−1018.

(38) Ludmann, K.; Ganea, C.; Váró, G. Back Photoreaction from Intermediate M of Bacteriorhodopsin Photocycle. *J. Photochem. Photobiol., B* **1999**, *49* (1), 23−28.

(39) Hendler, R. W.; Meuse, C. W.; Braiman, M. S.; Smith, P. D.; Kakareka, J. W. Infrared and Visible Absolute and Difference Spectra of Bacteriorhodopsin Photocycle Intermediates. *Appl. Spectrosc.* **2011**, *65* (9), 1029−1045.

(40) Rödig, C.; Chizhov, I.; Weidlich, O.; Siebert, F. Time-Resolved Step-Scan Fourier Transform Infrared Spectroscopy Reveals Differences between Early and Late M Intermediates of Bacteriorhodopsin. *Biophys. J.* **1999**, *76* (5), 2687−2701.

(41) Rothschild, K. J. FTIR difference spectroscopy of bacteriorhodopsin: toward a molecular model. *J. Bioenerg. Biomembr.* **1992**, *24* (2), 147−167.

(42) Korenstein, R.; Hess, B. Hydration Effects on the Photocycle of Bacteriorhodopsin in Thin Layers of Purple Membrane. *Nature* **1977**, *270* (5633), 184−186.

(43) Lórenz-Fonfría, V. A.; Furutani, Y.; Kandori, H. Active Internal Waters in the Bacteriorhodopsin Photocycle. A Comparative Study of the L and M Intermediates at Room and Cryogenic Temperatures by Infrared Spectroscopy. *Biochemistry* **2008**, *47* (13), 4071−4081.

(44) Dazzi, A.; Glotin, F.; Carminati, R. Theory of Infrared Nanospectroscopy by Photothermal Induced Resonance. *J. Appl. Phys.* **2010**, *107* (12), 124519.

(45) O'Callahan, B. T.; Yan, J.; Menges, F.; Muller, E. A.; Raschke, M. B. Photoinduced Tip−Sample Forces for Chemical Nanoimaging and Spectroscopy. *Nano Lett.* **2018**, *18* (9), 5499−5505.

(46) Lu, F.; Belkin, M. A. Infrared Absorption Nano-Spectroscopy Using Sample Photoexpansion Induced by Tunable Quantum Cascade Lasers. *Opt. Express* **2011**, *19* (21), 19942−19947.

(47) Mancini, A.; Giliberti, V.; Alabastri, A.; Calandrini, E.; De Angelis, F.; Garoli, D.; Ortolani, M. Thermoplasmonic Effect of Surface-Enhanced Infrared Absorption in Vertical Nanoantenna Arrays. *J. Phys. Chem. C* **2018**, *122* (24), 13072−13081.

(48) Baldassarre, L.; Giliberti, V.; Rosa, A.; Ortolani, M.; Bonamore, A.; Baiocco, P.; Kjoller, K.; Calvani, P.; Nucara, A. Mapping the Amide I Absorption in Single Bacteria and Mammalian Cells with Resonant Infrared Nanospectroscopy. *Nanotechnology* **2016**, *27* (7), 075101.

(49) Giliberti, V.; Badioli, M.; Baldassarre, L.; Nucara, A.; Calvani, P.; Ritter, E.; Puskar, L.; Hegemann, P.; Schade, U.; Ortolani, M. Vibrational Contrast Imaging and Nanospectroscopy of Single Cell Membranes by Mid-IR Resonantly-Enhanced Mechanical Photoexpansion. *2016 41st International Conference on Infrared, Millimeter, and Terahertz waves (IRMMW-THz)*, Copenhagen, Denmark, Sept. 25−30, 2016; pp 1−3.

(50) Müller, D. J.; Heymann, J. B.; Oesterhelt, F.; Möller, C.; Gaub, H.; Büldt, G.; Engel, A. Atomic force microscopy of native purple membrane. *Biochim. Biophys. Acta, Bioenerg.* **2000**, *1460* (1), 27−38.

(51) Marsh, D.; Pali, T. Infrared dichroism from the x-ray structure of bacteriorhodopsin. *Biophys. J.* **2001**, *80* (1), 305−312.

(52) Earnest, T. N.; Herzfeld, J.; Rothschild, K. J. Polarized Fourier Transform Infrared Spectroscopy of Bacteriorhodopsin. Transmembrane Alpha Helices Are Resistant to Hydrogen/Deuterium Exchange. *Biophys. J.* **1990**, *58* (6), 1539−1546.

(53) Earnest, T. N.; Roepe, P.; Braiman, M. S.; Gillespie, J.; Rothschild, K. J. Orientation of the Bacteriorhodopsin Chromophore Probed by Polarized Fourier Transform Infrared Difference Spectroscopy. *Biochemistry* **1986**, *25* (24), 7793−7798.

(54) Nabedryk, E.; Breton, J. Polarized Fourier Transform Infrared (FTIR) Difference Spectroscopy of the M 412 Intermediate in the Bacteriorhodopsin Photocycle. *FEBS Lett.* **1986**, *202* (2), 356−360.

(55) Amadei, A.; Daidone, I.; Di Nola, A.; Aschi, M. Theoretical-computational modelling of infrared spectra in peptides and proteins: a new frontier for combined theoretical-experimental investigations. *Curr. Opin. Struct. Biol.* **2010**, *20* (2), 155−61.

(56) Mukhopadhyay, S.; Cohen, S. R.; Marchak, D.; Friedman, N.; Pecht, I.; Sheves, M.; Cahen, D. Nanoscale electron transport and photodynamics enhancement in lipid-depleted bacteriorhodopsin monomers. *ACS Nano* **2014**, *8* (8), 7714−7722.

(57) Patil, A. V.; Premaruban, T.; Berthoumieu, O.; Watts, A.; Davis, J. J. Enhanced photocurrent in engineered bacteriorhodopsin monolayer. *J. Phys. Chem. B* **2012**, *116* (1), 683−689.

(58) Sasaki, J.; Shichidas, Y.; Lanyi, J. K.; Maeda, A. Protein Changes Associated with Reprotonation of the Schiff Base in the Photocycle of Asp96–>Asn Bacteriorhodopsin. The MN intermediate with unprotonated Schiff base but N-like protein structure. *J. Biol. Chem.* **1992**, *267* (29), 20782−20786.

(59) Zscherp, C.; Heberle, J. Infrared Difference Spectra of the Intermediates L, M, N, and O of the Bacteriorhodopsin Photoreaction Obtained by Time-Resolved Attenuated Total Reflection Spectroscopy. *J. Phys. Chem. B* **1997**, *101* (49), 10542−10547.

(60) Khatib, O.; Wood, J. D.; McLeod, A. S.; Goldflam, M. D.; Wagner, M.; Damhorst, G. L.; Koepke, J. C.; Doidge, G. P.; Rangarajan, A.; Bashir, R.; et al. Graphene-based platform for infrared near-field nanospectroscopy of water and biological materials in an aqueous environment. *ACS Nano* **2015**, *9* (8), 7968−7975.

(61) Oesterhelt, D.; Stoeckenius, W. Isolation of the cell membrane of Halobacterium halobium and its fractionation into red and purple membrane. *Methods Enzymol.* **1974**, *31*, 667−678.

(62) Blaudez, D.; Boucher, F.; Buffeteau, T.; Desbat, B.; Grandbois, M.; Salesse, C. Anisotropic optical constants of bacteriorhodopsin in the mid-infrared: consequence on the determination of alpha-helix orientation. *Appl. Spectrosc.* **1999**, *53* (10), 1299−1304.

(63) Nakano, T.; Kikugawa, G.; Ohara, T. A molecular dynamics study on heat conduction characteristics in DPPC lipid bilayer. *J. Chem. Phys.* **2010**, *133* (15), 154705.

(64) Fischer, H.; Polikarpov, I.; Craievich, A. F. Average protein density is a molecular-weight-dependent function. *Protein Sci.* **2004**, *13* (10), 2825−2828.

(65) Muddana, H. S.; Gullapalli, R. R.; Manias, E.; Butler, P. J. Atomistic simulation of lipid and DiI dynamics in membrane bilayers under tension. *Phys. Chem. Chem. Phys.* **2011**, *13* (4), 1368−1378.

(66) Morad, N. A.; Idrees, M.; Hasan, A. A. Specific heat capacities of pure triglycerides by heat-flux differential scanning calorimetry. *J. Therm. Anal.* **1995**, *45* (6), 1449−1461.

(67) Zander, U.; Bourenkov, G.; Popov, A. N.; De Sanctis, D.; Svensson, O.; McCarthy, A. A.; Round, E.; Gordeliy, V.; Mueller-Dieckmann, C.; Leonard, G. A. MeshAndCollect: an automated multi-crystal data-collection workflow for synchrotron macromolecular crystallography beamlines. *Acta Crystallogr., Sect. D: Biol. Crystallogr.* **2015**, *71* (11), 2328−2343.